%% file: manuscript.v3.tex
\documentclass[12pt]{article}

\usepackage{amsmath}
\usepackage{latexsym, amsbsy, amssymb,multirow, epsfig, amsmath,mathrsfs}
\usepackage{hyperref}
\usepackage{natbib,graphicx,setspace,lscape,longtable}
\usepackage{natbib,epsfig,graphicx,epstopdf}
\usepackage{amsmath,amsthm,amssymb,color}

\usepackage{makecell}
\def\beqr{\begin{eqnarray}}
\def\eeqr{\end{eqnarray}}
\def\beqrs{\begin{eqnarray*}}
\def\eeqrs{\end{eqnarray*}}

\def\eeqr{\end{eqnarray}}
\def\beqr{\begin{eqnarray}}
\def\eeqrs{\end{eqnarray*}}
\def\beqrs{\begin{eqnarray*}}

\def\be{\mbox{\boldmath$\eta$}}

\usepackage{latexsym}
\usepackage{epsfig}
\usepackage{bm}
\usepackage{makecell}
\usepackage{threeparttable}
\usepackage{dblfloatfix}
\usepackage{natbib,graphicx,setspace,lscape,longtable}
\usepackage{natbib,epsfig,graphicx}
\usepackage{amsmath,amsthm,amssymb,color}
\RequirePackage[mathlines, displaymath]{lineno}
\bibpunct{(}{)}{;}{a}{,}{,}
\usepackage{algorithm} 
\usepackage{algpseudocode} 
\usepackage{graphicx}
\usepackage{subcaption}
\usepackage{threeparttable}
\usepackage{dcolumn}
\usepackage{multirow}
\usepackage{booktabs}

\newtheorem{lem}{Lemma}

\input self-define-HL.tex

\input self-define-Heng.tex

\renewcommand{\log}{{\rm log}}

\title{ Distributed Iterative Hard Thresholding for Variable Selection in Tobit Models}
\author{Changxin Yang$^{a}$, Zhongyi Zhu$^{a}$ and \ Heng Lian$^{b}$
 \\
\begin{tabular}{l} 
{\small $^a$Department of Statistics, Fudan University, Shanghai, China}\\
{\small $^b$Department of Mathematics, City University of Hong Kong, Hong Kong, China}\\
\end{tabular}
}
\date{\today}
\providecommand{\keywords}[1]{\textbf{\textit{Keywords---}} #1}
\providecommand{\abstract}[1]{\textbf{\textit{Abstract---}} #1}
\newcounter{remark} 
\newtheorem{corollary}{Corollary}[section]

\begin{document}
\maketitle 

\abstract{While extensive research has been conducted on high-dimensional data and on regression with left-censored responses, simultaneously addressing these complexities remains challenging, with only a few proposed methods available. In this paper, we utilize the Iterative Hard Thresholding (IHT) algorithm on the Tobit model in such a setting. Theoretical analysis demonstrates that our estimator converges with a near-optimal minimax rate. Additionally, we extend the method to a distributed setting, requiring only a few rounds of communication while retaining the estimation rate of the centralized version. Simulation results show that the IHT algorithm for the Tobit model achieves superior accuracy in predictions and subset selection, with the distributed estimator closely matching that of the centralized estimator. When applied to high-dimensional left-censored HIV viral load data, our method also exhibits similar superiority.}

\section{Introduction}\label{sec:Introduction}
\par The analysis of left-censored data is a significant statistical focus and has attracted considerable research attention in recent years. It often arises due to the lower detection limit of an assay, posing a shared challenge across various fields such as biology, chemistry, and environmental sciences. For example, biological assays used to measure Human Immunodeficiency Virus (HIV) viral load in plasma may be limited in detecting concentrations below specific thresholds. The presence of such missingness renders commonly used linear regression methods ineffective. Furthermore, with advancements in modern data collection, these challenges often manifest in high-dimensional scenarios. In the context of HIV infection, there is a critical need to explore the association between the number of viral loads and extremely high-dimensional gene expression values. In addressing such challenges, various methodologies have been proposed in prior research. Among these, the Tobit model has proved to be valuable in modeling left-censored responses.

\par In recent years, significant progress has been made in the research on high dimensional censored data. A useful strategy to tackle high-dimensionality challenges involves constructing penalized estimators such as lasso-type estimators. \cite{muller2016} and  \cite{doi:10.1080/03610926.2014.904357} provide theoretical insights into the least absolute deviation estimator with the lasso penalty. \cite{soret2018lasso} propose the Lasso-regularized Buckley-James least square algorithm, extending the estimator in \cite{10.1093/biomet/66.3.429}. 
 \cite{doi:10.1080/07350015.2023.2182309} were the first to consider the high-dimensional Tobit model, which optimizes the likelihood function with a nonconvex penalty (specifically with the SCAD penalty in \cite{doi:10.1198/016214501753382273}).  In this paper, we concentrate on the IHT approach for variable selection. One advantage of using IHT is that the user can directly specify the number of variables to be retained, which may be useful in some scientific investigations.

\par IHT-style methods, which combine gradient descent with projection operations, have gained popularity in the literature for sparse recovery. Various algorithms have been proposed, such as standard IHT introduced in \cite{BLUMENSATH2009265}, GraDeS in \cite{Garg2009GradientDW}, and Hard Thresholding Pursuit (HTP) in \cite{doi:10.1137/100806278}. \cite{NIPS2014_218a0aef} demonstrated that the IHT procedure can achieve linear convergence to the optimal solution under the conditions of strong convexity and strong smoothness in high-dimensional settings. Extending this result,  \cite{WANG202336} considered nonsmooth loss functions under the less restrictive assumption of a locally positive-definite population Hessian. We will demonstrate that the linear convergence result also holds for the high-dimensional Tobit model.

\par Additionally, we consider a scenario where data are distributed across multiple locations and  develop
a distributed variant of our method. With advancements in data collection technology, gathering similar types of data from various regions has become increasingly common. Typically, directly aggregating all data at a central site faces practical hurdles related to storage, communication, and privacy concerns. To tackle these challenges, divide-and-conquer approaches are commonly employed. Many early approaches utilize one-shot methods, where estimators computed on local machines are transmitted to a central node and aggregated to form a global estimate, as in  \cite{JMLR:v14:zhang13b}, \cite{JMLR:v18:16-002}. However, such methods suffer from some drawbacks, as discussed by \cite{doi:10.1080/01621459.2018.1429274}, who proposed a communication-efficient distributed algorithm. Fortunately, the gradient descent-type update in the IHT method is conducive to integration with this algorithm. Building upon this concept, we introduce a communication-efficient estimator with an IHT-type update process. To our knowledge, this represents the first exploration of the high-dimensional Tobit model in a distributed setting. Our theoretical analysis and numerical results demonstrate that this estimator's convergence rate aligns with that achieved when pooling all data together while incurring low communication costs.
\par This article is structured as follows. In Section \ref{sec:2}, we review the Tobit model and then develop both  local (centralized) IHT and the distributed IHT. Section \ref{sec:convergence} is dedicated to our theoretical analysis. Section \ref{sec:6} draws some conclusions. All proofs are relegated to the Appendix.
\section{The Tobit model and IHT}\label{sec:2}

\subsection{Local IHT for Tobit model}
\par We consider the Tobit model, introduced in \cite{1da99cc8-515f-3e3f-9137-de96272a5db1}, a crucial tool for modeling a left-censored response. We assume the existence of a latent response variable \( y^* \) such that \( y=\max \left\{y^*, c_0\right\} \), where \( y^* \) follows a linear model 
$$ 
y^*=\mathbf{x}^{\prime} \boldsymbol{\beta}^*+\epsilon .
$$ 
Here, \( \mathbf{x}=\left(1, x_1, \ldots, x_d\right)^{\prime} \in \mathbb{R}^{d+1} \), \( \boldsymbol{\beta}^*=\left(\beta_0^*, \beta_1^*, \ldots, \beta_d^*\right)^{\prime} \in \mathbb{R}^{d+1} \), and \( \epsilon \sim \mathcal{N}\left(0, (\sigma^*)^2\right) \). We focus on a high-dimensional scenario where the dimension $d$ is much
larger than the sample size $n$, with \( \left\|\boldsymbol{\beta}^*\right\|_0=s_0 \). Without loss of generality, we assume \( c_0=0 \) throughout the following discussions.

Let $\{\mathbf{z}_i=\left( \mathbf{x}_i,y_i\right)\}_{i=1}^n$ represent an independent and identically distributed (i.i.d.) sample from the Tobit model, and define the indicator $d_i={I}_{y_i>0}$. The likelihood function for the censored response in the Tobit model can be expressed as
$$
\begin{aligned}
    L_n\left(\boldsymbol{\beta}, \sigma^2,\{\mathbf{z}_i\}_{i=1}^n\right)=\prod_{i=1}^n\left[\frac{1}{\sqrt{2 \pi} \sigma} \exp \left\{-\frac{1}{2 \sigma^2}\left(y_i-\mathbf{x}_i^{\prime} \boldsymbol{\beta}\right)^2\right\}\right]^{d_i}\left[\Phi\left(\frac{-\mathbf{x}_i^{\prime} \boldsymbol{\beta}}{\sigma}\right)\right]^{1-d_i},
\end{aligned}
$$
where $\Phi(\cdot)$ is the standard normal cumulative distribution function. While ${L}_n\left(\boldsymbol{\beta}, \sigma^2\right)$ is not concave in $\left(\boldsymbol{\beta}, \sigma^2\right)$, \cite{fd117be0-e2ff-33ae-997f-6ed0afb4cd80} found that the reparameterization $\boldsymbol{\delta}=\vbeta / \sigma$ and $\gamma^2=\sigma^{-2}$ results in a concave log-likelihood. Denoting $\boldsymbol{\theta}=(\boldsymbol{\delta},\gamma)$, after dropping ignorable constants, the negative log-likelihood is given by 
$$
\begin{aligned}
 \widehat{\mathcal{L}}(\boldsymbol{\theta})&= \frac{1}{n} \sum_{i=1}^n {\mathcal{L}}(\boldsymbol{\theta},\mathbf{z}_i) \\
&=\frac{1}{n} \sum_{i=1}^n d_i\left[-\log (\gamma)+\frac{1}{2}\left(\gamma y_i-\mathbf{x}_i^{\prime} \boldsymbol{\delta}\right)^2\right]-\left(1-d_i\right) \log \left(\Phi\left(-\mathbf{x}_i^{\prime} \boldsymbol{\delta}\right)\right). \\
\end{aligned}
$$
\par In the high-dimensional setting that $d$ is large, and maybe even larger than $n$, variable selection is desired. In this paper, we consider the hard constraint $\|\vdelta\|_0\le s$ where $s$ is a known upper bound of $s_0$. 
To optimize this constrained objective function, we use the IHT method, which is designed for high-dimensional regression. This method combines gradient descent with projection operations, making it computationally efficient. Furthermore, we slightly modify it for the Tobit model to prevent $\gamma$ from getting too small or even negative. Specifically, denoting the initial value as $\boldsymbol{\theta}^0$, for $t=0,1, \ldots$, our IHT algorithm can be formulated as
\begin{equation}
\boldsymbol{\theta}^{t+1}=P_{s,C^*}\left(\boldsymbol{\theta}^t-\eta \nabla \widehat{\mathcal{L}}\left(\boldsymbol{\theta}^t\right)\right) \doteq\left\{\begin{array}{l}
\boldsymbol{\delta}^{t+1}=P_s\left(\boldsymbol{\delta}^t-\eta \nabla_{\boldsymbol{\delta}} \widehat{\mathcal{L}}\left(\boldsymbol{\theta}^t\right)\right) \\
\gamma^{t+1}=\mathcal{T}_{C^*}(\gamma^t-\eta \nabla_\gamma \widehat{\mathcal{L}}\left(\boldsymbol{\theta}^t\right)) \label{ps}
\end{array}\right.,
\end{equation}
where $\eta>0$ is the step size, $P_s(\mathbf{\cdot})$ is the projection operator that retains only $s$ entries with the largest absolute values while setting other entries to zero, and
$$
\mathcal{T}_{C^*}(\gamma^{t}) =
\begin{cases}
    \gamma^t - \eta \nabla_\gamma \widehat{\mathcal{L}}\left(\boldsymbol{\theta}^t\right) & \text{if }  \gamma^t - \eta \nabla_\gamma \widehat{\mathcal{L}}\left(\boldsymbol{\theta}^t\right)\geq C^*\\
    C^* & \text{otherwise}
\end{cases}.
$$
\begin{algorithm}[H]
   \caption{ {Local IHT for Tobit Regression}} 
   \scalebox{0.8}{%
    \begin{minipage}{\linewidth}
    \begin{algorithmic}[1]
        \Require sparsity level $s$, number of iterations $T$,  step size $\eta$, lower bound $C^*$.
        \State Initialize $\boldsymbol{\theta}^0_{local}$
        \For{$t=0, \ldots, T-1$}
            \State Calculate the gradient $\nabla \widehat{\mathcal{L}}(\boldsymbol{\theta}^t_{local})$  
            \State  Update $\boldsymbol{\theta}^{t+1}_{local}$ as
            $$
            \boldsymbol{\theta}^{t+1}_{local}=P_{s,C^*}\left(\boldsymbol{\theta}^t_{local}-\eta \nabla \widehat{\mathcal{L}}\left(\boldsymbol{\theta}_{local}^t\right)\right)
            $$
        \EndFor
        \State \Return $\boldsymbol{\theta}^T_{local}=(\boldsymbol{\delta}^T_{local},\gamma^T_{local})$
    \end{algorithmic}
    \end{minipage}
    }
\end{algorithm}
Note that we have excluded $\gamma$ from the projection operation, as $\gamma$ is not part of the parameter for variable selection.  Additionally, we introduce a positive lower bound denoted as $C^*$ to prevent $\gamma\le 0$ during the update process.

\par In practical applications, the values of $s$ and $C^*$ are typically unknown and are treated as tuning parameters. 
There are two main approaches for determining $s$. First, the user can directly specify it based on a predefined desired sparsity, such as selecting the top 10 variables of interest. An alternative is to treat sparsity as a tuning parameter that can be optimized through cross-validation. $C^*$ can be set to a sufficiently small value to ensure it is smaller than $\gamma^*$. We present the entire IHT algorithm in Algorithm 1 (called local IHT to distinguish it from the distributed cases). Theorem \ref{sec:t1} in Section \ref{sec:convergence} provides a bound on the estimator error for the local IHT algorithm (after a sufficient number of iterations) as $O\left(\sqrt{\frac{s\log(n\vee d)}{n}}\right)$, which is well-known as the minimax rate for linear regression under the sparsity condition (if $s_0\ll d$, and $s$ and $s_0$ are of the same order). 

\subsection{ Distributed IHT for Tobit model}
\par In this section, we delve into the distributed implementation of the high-dimensional Tobit model across $M$ machines. This scenario holds significance in settings characterized by large volumes of training data. In such contexts, aggregating all raw data directly becomes impractical due to constraints like limited storage capacity, high communication costs, and privacy considerations. Hence, we are motivated to explore the development of a communication-efficient distributed IHT method.

\par For simplicity, we assume that each machine stores data of the same sample size $n$. Let $\mathcal{H}_m$ denote the subsample with a sample size of $n$ stored in the $m$-th machine $\mathcal{M}_m$, for $m = 1, \ldots, M$, and denote the global sample size $N = nM$. The local and global negative log-likelihood functions take the following form:
$$
    \begin{gathered}
\mathcal{L}_m(\boldsymbol{\theta})=\frac{1}{n} \sum_{i \in \mathcal{H}_m} {\mathcal{L}}(\boldsymbol{\theta},\mathbf{z}_i), \\
\mathcal{L}_N(\boldsymbol{\theta})=\frac{1}{M} \sum_{m=1}^M \mathcal{L}_m(\boldsymbol{\theta}).
\end{gathered}
$$
\par Our target is to find 
\begin{equation}{\boldsymbol{\theta}}^*=\underset{\boldsymbol{\theta}:\|\boldsymbol{\theta}\|_0 \leq s+1}{\operatorname{argmin}}   \mathcal{L}^*(\boldsymbol{\theta})  \doteq     E(\mathcal{L}_N(\boldsymbol{\theta})). \label{l*}
\end{equation}
 When direct aggregation of all data to optimize the global loss function is infeasible, a common approach is to approximate the global loss function via inter-machine communication. Subsequently, this approximation is used for optimization based on local data. \cite{doi:10.1080/01621459.2018.1429274} demonstrated that this approach could achieve optimal statistical precision with just a few rounds of communication. Following suit, we replace the original global negative log-likelihood function with an approximation. Specifically, given the current estimate $\overline{\boldsymbol{\theta}}$, we first collect the gradients $\nabla \mathcal{L}_m(\overline{\boldsymbol{\theta}})$ from different machines and then approximate the global negative log-likelihood function as
\begin{equation} 
    \widetilde{\mathcal{L}}(\boldsymbol{\theta}):=\mathcal{L}_1(\boldsymbol{\theta})-\left\langle\boldsymbol{\theta}, \nabla \mathcal{L}_1(\overline{\boldsymbol{\theta}})-\nabla \mathcal{L}_N(\overline{\boldsymbol{\theta}})\right\rangle \label{3},
\end{equation}
where $\mathcal{L}_1(\boldsymbol{\theta})$ is the local empirical negative log-likelihood function on the first machine $\mathcal{M}_1$ (treated as the central machine that can communicate with all others) and $\nabla \mathcal{L}_N(\overline{\boldsymbol{\theta}})=\frac{1}{M}\sum_{i=1}^M \nabla \mathcal{L}_m(\overline{\boldsymbol{\theta}})$. The global gradient for the approximate loss is $
\nabla \widetilde{\mathcal{L}}(\boldsymbol{\theta}):= \nabla \mathcal{L}_1(\boldsymbol{\theta})-(\nabla \mathcal{L}_1(\overline{\boldsymbol{\theta}})-\nabla \mathcal{L}_N(\overline{\boldsymbol{\theta}})).
$
Thus, based on the approximate global negative log-likelihood function, the IHT algorithm is, starting from $\vtheta^0=\overline\vtheta$,
 \begin{equation}
\boldsymbol{\theta}^{t+1}=P_{s,C^*}\left(\boldsymbol{\theta}^t- \eta \nabla \widetilde{\mathcal{L}}(\boldsymbol{\theta}^t)\right).\label{inner}
 \end{equation}
  After \eqref{inner} converges, ${\cal M}_1$ sends the new estimate to all other machines to obtain the updated approximate global negative log-likelihood function. Algorithm 2 presents the details of the distributed IHT algorithm for the Tobit model. Our theoretical results will show that the distributed IHT method can achieve the same statistical rate as the centralized version. Our simulation results further validate that it offers comparable statistical precision compared to its non-distributed counterpart.
 \begin{algorithm}[H]
    \caption{ {Distributed IHT for Tobit Regression}}
    \scalebox{0.7}{%
    \begin{minipage}{\linewidth}
    \begin{algorithmic}[1]
        \Require sparsity level $s$,  number of iterations in the inner loop $T$, number of iterations for the outer loop $Q$, step size $\eta$, lower bound $C^*$.
        \State Initialize $\boldsymbol{\theta}^0_{dis}$.
        \For{$q=0, \ldots, Q-1$}
        \State Let $\bar{\boldsymbol{\theta}}=\boldsymbol{\theta}^q_{dis}$ and ${\cal M}_1$ sends $\bar{\boldsymbol{\theta}}$ to all other machines.
        \State Each machine $\mathcal{M}_m$ computes $\nabla \mathcal{L}_m(\bar{\boldsymbol{\theta}})$ and send it to ${\cal M}_1$.
        \State $\mathcal{M}_1$ calculates $$ \nabla \mathcal{L}_N(\bar{\boldsymbol{\theta}})=\frac{1}{M} \sum_{m=1}^M \nabla \mathcal{L}_m(\bar{\boldsymbol{\theta}}).$$
        \State  On ${\cal M}_1$, let ${\boldsymbol{\theta}}^0_{inner}=\bar{\boldsymbol{\theta}}$ and execute the following inner loop.
        \For{$t=0, \ldots, T-1$}
        $$\boldsymbol{\theta}^{t+1}_{inner}=P_{s,C^*}\left(\boldsymbol{\theta}_{inner}^t- \eta \nabla \widetilde{\mathcal{L}}(\boldsymbol{\theta}_{inner}^t)\right).$$
        \EndFor
        \State Let $\boldsymbol{\theta}^{q+1}_{dis}=\boldsymbol{\theta}^T_{inner}$
        \EndFor
        \State \Return $\boldsymbol{\theta}^Q_{dis}=(\boldsymbol{\delta}^Q_{dis},\gamma^Q_{dis})$
    \end{algorithmic}
    \end{minipage}
    }
\end{algorithm}

\section{Convergence and statistical rates}\label{sec:convergence}
\par We present convergence guarantees for both the local IHT and the distributed IHT in this section. In the rest of the paper, $c$ and $C$ denote positive constants independent of $n$, $d$, $M$, whose values may vary across instances. In the following, for positive sequences $a_n$ and $b_n$, $a_n=O\left(b_n\right)$ or
 $a_n \lesssim b_n$ means $  a_n / b_n \leq c$ for some constant $c$, $a_n=o\left(b_n\right)$ 
  means $a_n/b_n\rightarrow 0 $, $a_n \asymp b_n$ means $a_n=O(b_n)$ and $b_n=O(a_n)$, $a_n\ll b_n$ or $b_n \gg a_n$ is the same as $a_n=o\left( b_n\right)$. 
We make the following assumptions.
\begin{itemize}
\item \textbf{Assumption 1 (Truncation and signal strength)}: $C^*$ is a constant that satisfies $\left.0<C^*<\gamma^*\right.$. $\|\vbeta^*\|_0=s_0$ and $\|\vbeta^*\|_2\le C$.

\item \textbf{Assumption 2 (Sparse eigenvalue condition)}: There exist positive constants $C_1$ and $\tilde{C}_2$ such that:
\begin{enumerate}
    \item[(i)]$$
    \begin{aligned}
    & C_1 \le E\left[\mathbf{a}'\begin{pmatrix} I_{y>0}\mathbf{x}\mathbf{x}' & -I_{y>0}\mathbf{x} y \\ -I_{y>0}y \mathbf{x}' & I_{y>0}y^2 \end{pmatrix}\mathbf{a}\right]<
    E\left[\mathbf{a}'\begin{pmatrix} \mathbf{x}\mathbf{x}' & -I_{y>0}\mathbf{x} y \\ -I_{y>0}y \mathbf{x}' & I_{y>0}y^2 \end{pmatrix} \mathbf{a}\right] \le \tilde{C}_2,
    \end{aligned}
    $$
  for all $(2s+s_0+1)$-sparse unit vectors $\mathbf{a}$.
    \item[(ii)] $\left(1 + \sqrt{\frac{s_0}{s}}\right)\frac{C_2-C_1}{C_2+C_1} < 1$ holds, where $C_2 \dot{=} \tilde{C}_2 + \frac{1}{\left(C^*\right)^2}$, and $s=O(s_0)$.
\end{enumerate}

\item \textbf{Assumption 3 (Sub-Gaussian designs)}: For some constant $C$, $\mathbf{x}_{i}$ are iid and $C$-sub-Gaussian, denoted as $\operatorname{subG}(C)$. That is, for all $\alpha \in \mathbb{R}^d$,
$$
\mathbb{E}[\exp(\alpha'\mathbf{x})] \leq \exp\left(\frac{C^2\|\alpha\|_2^2}{2}\right).
$$
\item \textbf{Assumption 4 (Proper initialization)}: $\|\boldsymbol{\theta}_{int}\|_0\leq (s+1)$ and $\gamma_{int}\geq C^*$.
\end{itemize}

\begin{remark}
To satisfy Assumption 1, we need to use a sufficiently small $C^*$. The sparse eigenvalue assumption is frequently used in high-dimensional data analysis. Note that here the assumption is imposed on the population quantity so even the standard eigenvalue assumption without constraining $\va$ to be sparse is reasonable. Examining the form of the population Hessian matrix $H(\boldsymbol{\theta})$ (seen in the Appendix), it is easy to see that Assumptions 1 and 2 (i) actually imply the Hessian matrix has eigenvalues between $C_1$ and $C_2$. 
Assumption 3 is also a standard assumption in high-dimensional regression analysis, making it possible to derive tight bounds on different quantities \citep{doi:10.1080/01621459.2018.1429274, belloni11}. Combined with the assumption $\|\vbeta^*\|_2\le C$ and the fact that the noise is Gaussian, it also implies $y_i$ is sub-Gaussian. Finally, a trivial initial value satisfying assumption 4 is $\vtheta_{int}=(\mathbf{0}, C^*)$.
\end{remark}

The following Theorem provides a bound for the $\ell_2$ estimation error of the local IHT algorithm, confirming the linear convergence of the IHT algorithm for the Tobit model.
\begin{theorem}\label{sec:t1}
(Local IHT for Tobit) Under Assumptions 1--4, choosing the step size $\eta=\frac{2}{C_1+C_2}$,
\begin{eqnarray*}
       \|\vtheta_{local}^{t+1}-\vtheta^*\|_2  
      &\leq &\left(\left(1+\sqrt{\frac{s_0}{s}}\right) \frac{C_2-C_1}{C_2+C_1}+C s\sqrt{\frac{\log(d\vee n)}{n}}\right) \|\vtheta_{local}^{t}-\vtheta^*\|_2\\
      &&+C\sqrt{s}\|\vtheta_{local}^{t}-\vtheta^*\|_2^2  +C \sqrt{\frac{s\log(d\vee n)}{n}},
\end{eqnarray*}
 with probability at least $1-(d \vee n)^{-C}$.
 In particular, if $
   \frac{s^2 \log (d \vee n)}{n}=o(1)$ and $\sqrt{s}\|\vtheta_{\text{int}}-\vtheta^*\|=o(1)$, after $T_1\asymp \log \left(\frac{n\left\|\boldsymbol{\theta}_{\text{int}}-\boldsymbol{\theta}^*\right\|^2_2}{s \log (d \vee n)}\right)$ iterations, we have
$$
\left\|\boldsymbol{\theta}_{\text{local}}^{T_1}-\boldsymbol{\theta}^*\right\|_2 \lesssim  \sqrt{\frac{s \log(d \vee n)}{n}},
$$
with probability at least $1-(d \vee n)^{-C}$.
\end{theorem}
The specific value chosen for $\eta$ is not the only option, and we opt for it for convenience in the proof. Our proof reveals that the requirement for $\eta$ and $s$ is actually that $\left(1+\sqrt{\frac{s_0}{s}}\right) \max \left\{\left|1-\eta C_1\right|,\left|1-\eta C_2\right|\right\}<1$. {For initialization, we recommend the Tobit-lasso estimator as it offers computational simplicity and has been shown to be consistent in \cite{doi:10.1080/07350015.2023.2182309}.}

\par Theorem \ref{sec:t1} establishes the convergece of the local IHT algorithm. The following Theorem concerns the convergence of the inner loop in the distributed setting, which is very similar to Theorem \ref{sec:t1}. This is not surprising since the approximate likelihood is expected to be close to the true likelihood. 

\begin{theorem}\label{sec:t2}
(Inner loop for Distributed IHT) Given $\bar{\boldsymbol{\theta}}=\boldsymbol{\theta}_{int}$ in Algorithm 2, under Assumptions 1--4, choosing the step size $\eta=\frac{2}{C_1+C_2}$, {if}  
$
   \frac{s^2 \log (d \vee n)}{n}=o(1)$ and $\sqrt{s}\|\bar\vtheta-\vtheta^*\|=o(1)$, then after $T_2 \asymp \log \left(\frac{n\|\bar\vtheta-\vtheta^*\|^2  }{s\log (d \vee n)}\right)$ iterations, we have
\begin{equation}
\left\|\boldsymbol{\theta}^{T_2}_{\text{inner}}-\boldsymbol{\theta}^*\right\|_2 \lesssim \sqrt{s}\left(\sqrt{\frac{ s\log(d\vee n)}{n}}\left\|\bar{\boldsymbol{\theta}}-\boldsymbol{\theta}^*\right\|_2+\|\bar{\boldsymbol{\theta}}-\boldsymbol{\theta}^*\|_2^2+\sqrt{\frac{ \log(d\vee n)}{N}}\right), \label{dis}
\end{equation} 
with probability at least $1-(d \vee n)^{-C}$.
\end{theorem}

The above result shows that after the inner loop ends, we have possibly a better estimate compared to when the loop starts. Combining such bounds over multiple stages $Q$, we get the following result.  

\begin{theorem}\label{cor:1}
 (Distributed IHT for Tobit) Under the assumptions of Theorem \ref{sec:t2}, after $Q \asymp \frac{\log N}{\log n}$ iterations in the outer loop (with the number of iterations in the inner loop as in the last statement of Theorem \ref{sec:t2}), with  probability at least $1-(d \vee n)^{-C}$, we have
$$
\left\|\boldsymbol{\theta}^Q_{dis}-\boldsymbol{\theta}^*\right\|_2 \lesssim \sqrt{\frac{s \log (d \vee n)}{N}}.
$$
\end{theorem}
\section{Conclusion}\label{sec:6}
\par In this paper, we delved into applying the IHT algorithm to the high-dimensional Tobit model and expanded its utility to distributed environments spanning multiple machines. This marks the first effort to study the Tobit model under the distributed setting in high-dimensional scenarios. We have shown that achieving a minimax convergence rate for local and distributed estimators is feasible. Simulation results validate our theoretical findings and demonstrate the good performance of our proposed method across diverse settings. When applied to high-dimensional left-censored HIV viral load data, the IHT method also effectively produced accurate predictions.

Regarding future directions, it is possible to explore extending the IHT algorithm to decentralized distributed algorithms, such as distributed gradient descent (DGD) as introduced in \cite{4749425}. DGD algorithms are extensively studied in the literature and offer advantages such as avoiding a single point of failure, addressing the limitations of centralized frameworks. Relaxing the requirement for sufficient local sample size by leveraging information from neighboring nodes in decentralized networks is also possible, as discussed in \cite{JMLR:v24:21-1333}. Exploring the combination with other distributed methods also remains an intriguing area for future research. Moreover, we can also consider a more complex class of missing data models known as Zero-Inflated models, where missing values are characterized as arising from a two-component mixture model. The presence of latent components in missing values requires a more careful analysis.
 
\medskip
\appendix
\section*{Appendix: Proofs}
\par The proof strategies for Theorem \ref{sec:t1} and Theorem \ref{sec:t2} share similarities. Therefore, we will first provide a detailed proof for Theorem \ref{sec:t2}, which is expected to be more challenging. 
We introduce some additional notations used in the proof. 
\begin{itemize}
    \item Let \(\phi(\cdot)\) and \(\Phi(\cdot)\) denote the standard normal probability density function (PDF) and cumulative distribution function (CDF), respectively. Let \(g(a) = \frac{\phi(a)}{\Phi(a)}\). Define \(h(a) = g(a)(a + g(a))=-g^\prime(a)\). According to \cite{10.1214/aoms/1177729093} and \cite{doi:10.1080/07350015.2023.2182309}, we have \(0 < h(a) < 1\) and \(-4.3 < h^\prime(a) < 0\).
\item Let \(\nabla \mathcal{L}_1^t = \nabla \mathcal{L}_1\left(\boldsymbol{\theta}^t\right)\) denote the negative log-likelihood of the first machine, and similarly for \(\nabla\tilde\calL^t=\nabla\tilde{ \mathcal{L}}(\boldsymbol{\theta}^t)\). 
 The gradient of the log-likelihood \(\nabla \mathcal{L}_1(\boldsymbol{\theta})\) can be expressed as 
\begin{equation}
     \frac{1}{n} \sum_{i\in\calH_1}\left[\begin{array}{c}
-I_{y_i>0} \mathbf{x}_i\left(\gamma y_i-\mathbf{x}_i^{\prime} \vdelta\right)+I_{y_i \leq 0}\left(\mathbf{x}_i^{\prime} g\left(-\mathbf{x}^{\prime}_i \boldsymbol{\delta}\right)\right) \\
-I_{y_i>0}\left(\gamma^{-1}-y_i^{\prime}\left(\gamma y_i-\mathbf{x}_i^{\prime} \vdelta\right)\right)
\end{array}\right]. \label{gr}
\end{equation}
\item Let \(H(\boldsymbol{\theta})\) denote the population Hessian matrix, given by
\begin{equation}
    E\left(\begin{bmatrix}
I_{y>0} \mathbf{x x}^{\prime} + I_{y\leq 0} \mathbf{x x}^{\prime} h\left(-\mathbf{x}^{\prime} \boldsymbol{\delta}\right) & -I_{y>0} \mathbf{x} y \\
-I_{y>0} y \mathbf{x}^{\prime} & I_{y>0} (y^2+\gamma^{-2}) \
\end{bmatrix}\right). \label{hm}
\end{equation}

\item Define \(I^t = S^t \cup S_0\), where \(S^t\) and \(S_0\) represent the support (indices of nonzero entries) of \(\boldsymbol{\theta}^t\) and the true \(\boldsymbol{\theta}^*\), respectively.

\item Given a vector \(\boldsymbol{\theta}\), a matrix \(B\), and an index set \(I \subseteq \{1, \ldots, d+2\}\), \(\boldsymbol{\theta}_I\) denotes the vector obtained from \(\boldsymbol{\theta}\) by setting components not in \(I\) to zero, while \(B_{I}\) represents the matrix obtained from \(B\) by setting the components of rows not in \(I\) to zero. \(|I|\) represents the cardinality of the set \(I\).

\item We say that a random variable \(X\) is sub-exponential with parameters \(C\) if the $L^p$ norm of \(X\) satisfy
\[
\|X\|_{L^p}=\left({E}|X|^p\right)^{1 / p} \leq C p, \quad \text { for all } p \geq 1.
\]
\item Define \(\mathbf{x}_{\max}=\max_{i\in\{1,\dots,n\}} ||\mathbf{x}_i||_\infty\), and \(y_{\max}=\max_{i\in\{1,\dots,n\}}|y_i|\).
\end{itemize}

\subsection*{Proof of Theorem \ref{sec:t2}}
\par 
In the inner loop, we have
\[
\begin{aligned}
 e_{t+1} & \doteq \left\|\boldsymbol{\theta}^{t+1}-\boldsymbol{\theta}^*\right\|_2 \\
& =\left\|\left(P_{s,C^*}\left(\boldsymbol{\theta}^t-\eta \nabla\tilde{ \mathcal{L}}^t\right)\right)_{I^{t+1}}-\boldsymbol{\theta}^*_{I^{t+1}}\right\|_2 \\
& \leq \left\|\left(P_{s}\left(\boldsymbol{\theta}^t-\eta \nabla\tilde{ \mathcal{L}}^t\right)\right)_{I^{t+1}}-\left(\boldsymbol{\theta}^t-\eta \nabla\tilde{ \mathcal{L}}^t\right)_{I^{t+1}}\right\|_2 \\
& \quad + \left\|\left(\boldsymbol{\theta}^t-\eta \nabla\tilde{ \mathcal{L}}^t\right)_{I^{t+1}}-\boldsymbol{\theta}^*_{I^{t+1}}\right\|_2 \\
& \leq \left(1+\sqrt{\frac{\left|I^{t+1}\right|-s}{\left|I^{t+1}\right|-s_0}}\right)\left\|\left(\boldsymbol{\theta}^t-\eta \nabla\tilde{ \mathcal{L}}^t\right)_{I^{t+1}}-\boldsymbol{\theta}^*_{I^ {t+1}}\right\|_2.
\end{aligned}
\]
The first inequality above holds because, under Assumption 1, if \(\gamma^t - \eta \nabla_\gamma \tilde{\mathcal{L}}\left(\boldsymbol{\theta}^t\right) < C^*\), then it is evident that \(\left|\gamma^*-C^*\right| \leq \left|\gamma^*-\gamma^t-\eta \nabla_\gamma \tilde{\mathcal{L}}\left(\boldsymbol{\theta}^t\right)\right|\), meaning removing the thresholding with $C^*$ can only increase the error. The 2nd inequality is due to Lemma 1 in \cite{NIPS2014_218a0aef}. 

\par Then, by adding and subtracting terms, we have
\[ 
\begin{aligned}
 & \left\|\left({\boldsymbol{\theta}}^t-\eta \nabla\tilde{ \mathcal{L}}^t\right)_{I^{t+1}}-\boldsymbol{\theta}^*_{I^ {t+1}}\right\|_2 \\
& \leq \left\|{\boldsymbol{\theta}}_{I^{t+1}}^t-\boldsymbol{\theta}^*_{I^{t+1}}-\eta E\left(\nabla\tilde{ \mathcal{L}}^t-\nabla\tilde{ \mathcal{L}}\left(\boldsymbol{\theta}^*\right)\right)_{I^{t+1}}\right\|_2 \\
& \quad + \eta \left\|E\left(\nabla\tilde{ \mathcal{L}}^t-\nabla\tilde{ \mathcal{L}}\left(\boldsymbol{\theta}^*\right)\right)_{I^{t+1}}-\left(\nabla\tilde{ \mathcal{L}}^t-\nabla\tilde{ \mathcal{L}}\left(\boldsymbol{\theta}^*\right)\right)_{I^{t+1}}\right\|_2 \\
& \quad + \eta \left\|\left(\nabla\tilde{ \mathcal{L}}\left(\boldsymbol{\theta}^*\right)\right)_{I^{t+1}}\right\|_2 \\
&= \left\|{\boldsymbol{\theta}}_{I^{t+1}}^t-\boldsymbol{\theta}^*_{I^{t+1}}-\eta E\left(\nabla \mathcal{L}_1^t-\nabla{ \mathcal{L}_1}\left(\boldsymbol{\theta}^*\right)\right)_{I^{t+1}}\right\|_2 \\
& \quad + \eta \left\|E\left(\nabla \mathcal{L}_1^t-\nabla{ \mathcal{L}_1}\left(\boldsymbol{\theta}^*\right)\right)_{I^{t+1}}-\left(\nabla\mathcal{L}_1^t-\nabla{ \mathcal{L}_1}\left(\boldsymbol{\theta}^*\right)\right)_{I^{t+1}}\right\|_2 \\
& \quad + \eta \left\|\left(\nabla\tilde{ \mathcal{L}}\left(\boldsymbol{\theta}^*\right)\right)_{I^{t+1}}\right\|_2 \\
& \doteq ||P_1||_2+\eta ||P_2||_2+\eta ||P_3||_2.
\end{aligned}
\]
\par For $P_1$, by Assumption 2 and Taylor's expansion, we get
\begin{equation}
      \begin{aligned}
& \left\|\boldsymbol{\theta}_{I^{t+1}}^t-\boldsymbol{\theta}^*_{I^{t+1}}-\eta\left(E \nabla \mathcal{L}_1^t-E \nabla \mathcal{L}_1\left(\boldsymbol{\theta}^*\right)\right)_{I^{t+1}}\right\|_2 \\
& \leq \|(I-\eta H(\boldsymbol{\alpha}))_{I^{t+1}\cup S^t}\left(\boldsymbol{\theta}_{I^{t+1}\cup S^t}^t-\boldsymbol{\theta}^*_{I^{t+1}\cup S^t}\right) \|_2 \\
& \leq \max \left\{\left|1-\eta C_1\right|,\left|1-\eta C_2\right|\right\}\left\|\boldsymbol{\theta}^t-\boldsymbol{\theta}^*\right\|_2,
\end{aligned} \label{p1}
\end{equation}

where \( \boldsymbol{\alpha} \) lies on the line between \( \boldsymbol{\theta}^t \) and \( \boldsymbol{\theta}^* \). 
\par For $P_2$, by Lemma \ref{lem:grad}, with  probability at least $1-(d \vee n)^{-C}$, we have

\begin{equation}
\begin{aligned}
 \|P_2\|_{\infty} 
& \lesssim \sqrt{\frac{s \log (d \vee n)}{n}}  e_t+e_t^2,
\end{aligned} \label{17}
\end{equation}
and thus $\|P_2\|_2\lesssim  s\sqrt{\frac{  \log (d \vee n)}{n}}  e_t +\sqrt{s}e_t^2$.

\par  {For $P_3$}, by Lemma \ref{lem:2}, we have, with probability at least $1-(d \vee n)^{-C}$,
$$
\begin{aligned}
     \| P_3 \|_2 &\lesssim    s\sqrt{\frac{  \log (d \vee n)}{n}}  e_0  +\sqrt{s}  e_0^2+\sqrt{\frac{s \log (d \vee n)}{N}}.
\end{aligned}
$$

Therefore, by choosing \(\eta = \frac{2}{C_2+C_1}\) and combining bounds for \(P_1\), \(P_2\), and \(P_3\), we have
\bse
e_{t+1} 
& \leq&\left(\left(1+\sqrt{\frac{s_0}{s}}\right) \frac{C_2-C_1}{C_2+C_1}+Cs\sqrt{\frac{ \log (d \vee n)}{n}}\right) e_t+\sqrt{s}e_t^2\\
&&+Cs\sqrt{\frac{ \log (d \vee n)}{n}}  e_0+C\sqrt{s}e_0^2+\sqrt{\frac{s \log (d \vee n)}{N}},
\ese
with probability at least \(1-(d \vee n)^{-C}\).

Thus, if $s\sqrt{\frac{ \log (d \vee n)}{n}}=o(1)$ and $\sqrt{s}e_0=o(1)$, after \(T \asymp \left( \log \left(\frac{ne_0}{s \log (d \vee n)}\right)\right)\) iterations, with probability at least \(1-(d \vee n)^{-C}\),  
\begin{equation}
    e_T \lesssim \sqrt{s}\left(\sqrt{\frac{ s\log(d\vee n)}{n}}\left\|\bar{\boldsymbol{\theta}}-\boldsymbol{\theta}^*\right\|_2+ \|\bar{\boldsymbol{\theta}}-\boldsymbol{\theta}\|_2^2+\sqrt{\frac{ \log(d\vee n)}{N}}\right). \label{fp}
\end{equation}
  The proof is completed.

\subsection*{Proof of Theorem \ref{sec:t1}}
\par In fact, the proof of Theorem \ref{sec:t1} is essentially the same as the proof of Theorem \ref{sec:t2}.
Our goal is to provide an upper bound for
\[
\begin{aligned}
 e_{t+1} & \doteq \left\|\boldsymbol{\theta}^{t+1}-\boldsymbol{\theta}^*\right\|_2   \leq \left(1+\sqrt{\frac{\left|I^{t+1}\right|-s}{\left|I^{t+1}\right|-s_0}}\right)\left\|\left(\boldsymbol{\theta}^t-\eta \nabla\widehat{ \mathcal{L}}^t\right)_{I^{t+1}}-\boldsymbol{\theta}^*_{I^ {t+1}}\right\|_2.
\end{aligned}
\]
We have
\[
\begin{aligned}
 & \left\|\left({\boldsymbol{\theta}}^t-\eta \nabla\widehat{ \mathcal{L}}^t\right)_{I^{t+1}}-\boldsymbol{\theta}^*_{I^ {t+1}}\right\|_2 \\
& \leq \left\|{\boldsymbol{\theta}}_{I^{t+1}}^t-\boldsymbol{\theta}^*_{I^{t+1}}-\eta E\left(\nabla\widehat{ \mathcal{L}}^t-\nabla\widehat{ \mathcal{L}}\left(\boldsymbol{\theta}^*\right)\right)_{I^{t+1}}\right\|_2 \\
& \quad + \eta \left\|E\left(\nabla\widehat{ \mathcal{L}}^t-\nabla\widehat{ \mathcal{L}}\left(\boldsymbol{\theta}^*\right)\right)_{I^{t+1}}-\left(\nabla\widehat{ \mathcal{L}}^t-\nabla\widehat{ \mathcal{L}}\left(\boldsymbol{\theta}^*\right)\right)_{I^{t+1}}\right\|_2 \\
& \quad + \eta \left\|\left(\nabla\widehat{ \mathcal{L}}\left(\boldsymbol{\theta}^*\right)\right)_{I^{t+1}}\right\|_2 \\
& \doteq ||P_1||_2+\eta ||P_2||_2+\eta ||P_3||_2.
\end{aligned}
\]
\par For \(P_1\), by Assumption 2 and Taylor's expansion, we can obtain that
\[
\left\|P_1\right\|_2 \leq  \max \left\{\left|1-\eta C_1\right|,\left|1-\eta C_2\right|\right\}e_t.
\]

\par {For $P_2$}, similar to \eqref{17}, with probability at least $1-(d \vee n)^{-C}$, we have
$$
\begin{aligned}
\left\|P_2\right\|_2 & \lesssim s\sqrt{\frac{\log(d\vee n)}{n}} e_t+\sqrt{s}e_t^2.
\end{aligned}
$$
 \par For $P_3$, as we have shown in the proof of Lemma \ref{lem:2},  $\left\|\nabla \widehat{\mathcal{L}}\left(\boldsymbol{\theta}^*\right)\right\|_{2}\lesssim \sqrt{\frac{s\log(d\vee n)}{n}}$ with probability at least $1-(d \vee n)^{-C}$.

Combining the bounds and choosing $\eta = \frac{2}{C_2+C_1}$, we get
\[
\begin{aligned}
       e_{t+1} & 
      &\leq \left(\left(1+\sqrt{\frac{s_0}{s}}\right) \frac{C_2-C_1}{C_2+C_1}+C s\sqrt{\frac{\log(d\vee n)}{n}}\right) e_t+C\sqrt{s}e_t^2+C \sqrt{\frac{s\log(d\vee n)}{n}},
\end{aligned}
\]
with probability of at least \(1-(d \vee n)^{-C}\).

\par Thus if $s\sqrt{\frac{\log(d\vee n)}{n}}=o(1)$ and $\sqrt{s}e_0=o(1)$, after $T_1\asymp \log \left(\frac{n\left\|\boldsymbol{\theta}^0-\boldsymbol{\theta}^*\right\|^2_2}{s \log (d \vee n)}\right) $ iterations , with probability at least $1-(d \vee n)^{-C}$, we have 
$$
e_T \lesssim \sqrt{\frac{ s\log(d\vee n)}{n}}.
$$
 
\subsection*{Proof of Theorem \ref{cor:1}.}
 Theorem \ref{sec:t2} implies
\[
\begin{aligned}
    \left\|\boldsymbol{\theta}_{dis}^1-\boldsymbol{\theta}^*\right\|_2  
    & \lesssim \left(s \sqrt{\frac{\log(d\vee n)}{n}}\left\|{\boldsymbol{\theta}_{dis}^0}-\boldsymbol{\theta}^*\right\|_2+\sqrt{s}\left\|{\boldsymbol{\theta}_{dis}^0}-\boldsymbol{\theta}^*\right\|_2^2+\sqrt{\frac{ s\log(d\vee n)}{N}}\right).
\end{aligned} 
\]

This implies that after $Q\asymp \frac{\log N}{\log n}$ iterations for the outer loop (note if $n\ge N^c$ for some constant $c\in (0,1)$, $Q$ is a constant) we obtain
\[
\left\|\boldsymbol{\theta}_{dis}^Q-\boldsymbol{\theta}^*\right\|_2 \lesssim \sqrt{\frac{s \log (d \vee n)}{N}}.
\]

\subsection*{Technical Lemmas}
\begin{lem}\label{lem:1}
Given $\boldsymbol{\theta}=(\boldsymbol{\delta},\gamma)$ satisfying $||\boldsymbol{\theta}-\boldsymbol{\theta}^*||_0\leq (2s+s_0+1)$ and $\gamma\geq C^*$, under Assumptions 1, 2 and 3, then, with probability at least $1-(d \vee n)^{-C}$, 
$$
\left\|\left(\nabla^2 \mathcal{L}_1\left({
\boldsymbol{\theta}
}\right)-\nabla^2 \mathcal{L}_1\left(\boldsymbol{\theta}^*\right)\right) (\boldsymbol{\theta}-\boldsymbol{\theta}^*)\right\|_{\infty} \lesssim \|{
\boldsymbol{\theta}}-\boldsymbol{\theta}^*\|_2^2.
$$
\end{lem}

\par \noindent\textbf{Proof.} Denote $\boldsymbol{\theta}-\boldsymbol{\theta}^*=\boldsymbol{\Delta}=\left(\boldsymbol{\Delta}_{\boldsymbol{\delta}}, \boldsymbol{\Delta}_\gamma\right)$. 
\begin{itemize}
    \item For $j=1,\dots,d+1$, 
    $$
\begin{aligned}
& {\left[\left(\nabla^2 \mathcal{L}_1\left(\boldsymbol{\theta}^*+\boldsymbol{\Delta}\right)-\nabla^2 \mathcal{L}_1\left(\boldsymbol{\theta}^*\right)\right) \boldsymbol{\Delta}\right]} _j \\
& =\frac{1}{n} \sum_{i=1}^n\left(1-d_i\right)\left(x_{i j} \mathbf{x}_i^{\prime} \boldsymbol{\Delta}_{\boldsymbol{\delta}}\left(h\left(\mathbf{x}_i^{\prime} \boldsymbol{\delta}^*\right)-h\left(\mathbf{x}_i^{\prime} \boldsymbol{\delta}\right)\right)\right. \\
&\le \frac{1}{n}\sum_i|x_{ij}|(\vx_i'\vDelta_{\boldsymbol{\delta}})^2 
  \lesssim \|\boldsymbol{\delta}\|_2^2,
\end{aligned}
    $$
 where we use $|h'(a)|<4.3$ for all $a \in \mathbb{R}$ (\cite{doi:10.1080/07350015.2023.2182309}).
    \item For $j=d+2$,
    $$
   \begin{aligned}
{\left[\left(\nabla^2 \mathcal{L}_1\left(\boldsymbol{\theta}^*+\boldsymbol{\Delta}\right)-\nabla^2 \mathcal{L}_1\left(\boldsymbol{\theta}^*\right)\right) \boldsymbol{\Delta}\right]_j } & =\frac{1}{n} \sum_{i=1}^n d_i\left((\gamma)^{-2}-\left(\gamma^*\right)^{-2}\right) \boldsymbol{\Delta}_\gamma \\
& \leq \frac{2}{\left(C^*\right)^3} \boldsymbol{\Delta}_\gamma^2,
\end{aligned}
    $$
     where the last inequality holds because of $\gamma\geq C^*$.
\end{itemize}
 Thus, with probability of at least $1 - (d \vee n)^{-C}$,
$$
\begin{aligned}
\left\|\left(\nabla^2 \mathcal{L}_1(\boldsymbol{\theta})-\nabla^2 \mathcal{L}_1\left(\boldsymbol{\theta}^*\right)\right) \boldsymbol{\Delta}\right\|_{\infty} 
& \lesssim \|\boldsymbol{\Delta}\|_2^2.
\end{aligned}
$$

\begin{lem}\label{lem:grad}
For $ {\boldsymbol{\theta}}=( {\boldsymbol{\delta}}, {\gamma})$ satisfying $|| {\boldsymbol{\theta}}-\boldsymbol{\theta}^*||_0\leq (2s+s_0+1)$ and $ {\gamma}\geq C^*$, under Assumptions 1, 2, and 3, with probability at least $1-(d \vee n)^{-C}$, we have with probability $1-(d\vee n)^{-C}$,
\bse
\|\nabla \calL_1(\vtheta)-\nabla\calL_1(\vtheta^*)-E[\calL_1(\vtheta)-\nabla\calL_1(\vtheta^*)] \|_\infty\le C\sqrt{\frac{s\log(d\vee n)}{n}}\|\vtheta-\vtheta^*\|_2+C\|\vtheta-\vtheta^*\|^2_2.
\ese
\end{lem}
\par \noindent\textbf{Proof.} 
We have, by Taylor's expansion,
\bse
 &&\left(\nabla \mathcal{L}_1( {\boldsymbol{\theta}})-\nabla \mathcal{L}_1\left(\boldsymbol{\theta}^*\right)\right)  \\
 &=&\nabla^2 \mathcal{L}_1\left(\boldsymbol{\theta}^*\right)\left( {\boldsymbol{\theta}}-\boldsymbol{\theta}^*\right)   +\int_{0}^{1} \left(\nabla^2 \mathcal{L}_1\left(\boldsymbol{\theta}^*+u\left( {\boldsymbol{\theta}}-\boldsymbol{\theta}^*\right)\right)-\nabla^2 \mathcal{L}_1\left(\boldsymbol{\theta}^*\right)\right)\left( {\boldsymbol{\theta}}-\boldsymbol{\theta}^*\right)du .
\ese
By Lemma \ref{lem:1}, the second term above as well as its expectation (can be shown following exactly the same lines) is bounded by $C\|\vtheta-\vtheta^*\|_2^2$.
For the 1st term above, we first derive a bound for
\begin{equation}
    \begin{aligned}
& \left\|\nabla^2 \mathcal{L}_1\left(\boldsymbol{\theta}^*\right)-\nabla^2 \mathcal{L}^*\left(\boldsymbol{\theta}^*\right)\right\|_{\max},
    \end{aligned}  \label{19}
\end{equation}
where we denote $\|H\|_{\max}\doteq \max{|H_{ij}|}$ for a matrix $H$ and we recall that by our notation $\mathcal{L}^*(\boldsymbol{\theta}^*)=E[\mathcal{L}_1 (\boldsymbol{\theta}^*)]$. 
  We analyze each entry of the empirical Hessian matrix $\nabla^2\calL_1(\vtheta^*)$. 
\begin{itemize}
    \item 
    $
\frac{\partial^2}{\partial \boldsymbol{\delta}_j \partial \boldsymbol{\delta}_k} \mathcal{L}_1(\boldsymbol{\theta})=\frac{1}{n}\sum_{i=1}^n x_{i j} x_{i k}\left[d_i+\left(1-d_i\right) h\left(-\mathbf{x}_i^{\prime} \boldsymbol{\delta}\right)\right] .
$
Since $0<h(s)<1$ for all $s \in \mathbb{R}$, we see that  $$|x_{i j} x_{i k}\left[d_i+\left(1-d_i\right) h\left(-\mathbf{x}_i^{\prime} \boldsymbol{\delta}\right)\right]|\leq |x_{i j} x_{i k}|.$$ 
Thus, $x_{i j} x_{i k}\left[d_i+\left(1-d_i\right) h\left(-\mathbf{x}_i^{\prime} \boldsymbol{\delta}\right)\right]$ is sub-exponential.
\item $\frac{\partial^2}{\partial \boldsymbol{\delta}_j \partial \gamma} \mathcal{L}_1(\boldsymbol{\theta})=\frac{1}{n}\sum_{i=1}^n -d_i y_i x_{i j}$. It is easy to see that $d_iy_i x_{i j}$ is sub-exponential.
\item $\frac{\partial^2}{\partial \gamma^2} \mathcal{L}_1(\boldsymbol{\theta})=\frac{1}{n}\sum_{i=1}^n d_i\left(\gamma^{-2}+y_i^{ 2}\right)$, and $d_i\left(\gamma^{-2}+y_i^{ 2}\right)$ is again sub-exponential.
\end{itemize}
With the aid of the concentration inequality for sub-exponential variables, \eqref{19} is bounded by $C\sqrt{\frac{\log(d\vee n)}{n}}$ with probability at least \(1-(d \vee n)^{-C}\). Consequently,
\[
\left\|(\nabla^2 \mathcal{L}_1\left(\boldsymbol{\theta}^*\right)-\nabla^2 \mathcal{L}^*\left(\boldsymbol{\theta}^*\right))({\boldsymbol{\theta}}-\boldsymbol{\theta}^*)\right\|_{\infty}\leq C\sqrt{\frac{\log(d\vee n)}{n}}||{\boldsymbol{\theta}}-\boldsymbol{\theta}^*||_1\le C\sqrt{\frac{s\log(d\vee n)}{n}}\|{\boldsymbol{\theta}}-\boldsymbol{\theta}^*\|_2.
\]
This completes the proof.

\begin{lem}\label{lem:2} Given $\bar{\boldsymbol{\theta}}=(\bar{\boldsymbol{\delta}},\bar{\gamma})$ satisfying $||\bar{\boldsymbol{\theta}}-\boldsymbol{\theta}^*||_0\leq (2s+s_0+1)$ and $\bar{\gamma}\geq C^*$,
under Assumptions 1, 2, and 3, with probability at least $1-(d \vee n)^{-C}$, we have,
$$
\begin{aligned}
\left\|\nabla \widetilde{\mathcal{L}}\left(\boldsymbol{\theta}^*\right)\right\|_{\infty} & \lesssim 
\sqrt{\frac{s\log (d \vee n)}{n}}\left\|\overline{\boldsymbol{\theta}}-\boldsymbol{\theta}^*\right\|_2+ \|\overline{\boldsymbol{\theta}}-\boldsymbol{\theta}\|_2^2+\sqrt{\frac{\log (d \vee n)}{N}}. 
\end{aligned}
$$
\end{lem}
\par \noindent\textbf{Proof.} 
$$
    \begin{aligned}
\nabla \tilde{\mathcal{L}}\left(\boldsymbol{\theta}^*\right)  =&\nabla \mathcal{L}_1\left(\boldsymbol{\theta}^*\right)-\nabla \mathcal{L}_1(\bar{\boldsymbol{\theta}})+\mathcal{L}_N(\bar{\boldsymbol{\theta}}) \\
 =&\left(\nabla \mathcal{L}_N(\bar{\boldsymbol{\theta}})-\nabla \mathcal{L}_N\left(\boldsymbol{\theta}^*\right)\right)-\left(\nabla \mathcal{L}_1(\bar{\boldsymbol{\theta}})-\nabla \mathcal{L}_1\left(\boldsymbol{\theta}^*\right)\right)+\nabla \mathcal{L}_N\left(\boldsymbol{\theta}^*\right).
\end{aligned}
$$
By Lemma \ref{lem:grad}, with probability at least $1-(d \vee n)^{-C}$, we have
$$
\begin{aligned}
\left\|\nabla \widetilde{\mathcal{L}}\left(\boldsymbol{\theta}^*\right)\right\|_{\infty}  \lesssim &\sqrt{\frac{s\log (d \vee n)}{n}}\left\|\overline{\boldsymbol{\theta}}-\boldsymbol{\theta}^*\right\|_2+ \|\overline{\boldsymbol{\theta}}-\boldsymbol{\theta}\|_2^2+\left\|\nabla \mathcal{L}_N\left(\boldsymbol{\theta}^*\right)\right\|_{\infty} .
\end{aligned}
$$

Recall that $0 < |g^\prime(s)| < 1$, and thus $|g(-\mathbf{x}^{\prime} \boldsymbol{\delta})-g(0)|\le |\vx'\vdelta|$, implying $g(-\mathbf{x}^{\prime} \boldsymbol{\delta})$ is sub-Gaussian. Thus using the same arguments as in the proof of Lemma \ref{lem:grad}, it is easy to see, based on the expression \eqref{gr} for the gradient, that the components of $\mathcal{L}_N(\boldsymbol{\theta}^*)$ are all sub-exponential. Thus $\left\|\nabla \mathcal{L}_N\left(\boldsymbol{\theta}^*\right)\right\|_{\infty} \lesssim \sqrt{\frac{\log(d\vee n)}{N}}$ holds with probability at least $1-(d \vee n)^{-C}$.

\bibliographystyle{ECA_jasa}
\bibliography{papers,books}
\end{document}

%% file: self-define-HL.tex
\def\log{\hbox{log}}

\def\boxit#1{\vbox{\hrule\hbox{\vrule\kern6pt
          \vbox{\kern6pt#1\kern6pt}\kern6pt\vrule}\hrule}}

\def\bse{\begin{eqnarray*}}
\def\ese{\end{eqnarray*}}
\def\be{\begin{eqnarray}}
\def\ee{\end{eqnarray}}
\def\bq{\begin{equation}}
\def\eq{\end{equation}}

\newtheorem{proposition}{Proposition}

\newcommand{\blem}{\begin{lemma}}
\newcommand{\elem}{\end{lemma}}
\newcommand{\bthe}{\begin{theorem}}
\newcommand{\ethe}{\end{theorem}}

\newtheorem{definition}{Definition}
\newtheorem{lemma}[definition]{Lemma}
\newtheorem{theorem}[definition]{Theorem}

\def\delete#1{\iffalse #1 \fi}

\def\bse{\begin{eqnarray*}}
\def\ese{\end{eqnarray*}}
\def\bee{\begin{enumerate}}
\def\eee{\end{enumerate}}
\def\bqe{\begin{eqnarray}}
\def\eqe{\end{eqnarray}}
\def\bed{\begin{description}}
\def\eed{\end{description}}
\def\bei{\begin{itemize}}
\def\eei{\end{itemize}}

\def\pmb#1{\setbox0=\hbox{#1}%
    \kern-.025em\copy0\kern-\wd0
    \kern.05em\copy0\kern-\wd0
    \kern-.025em\raise.0433em\box0 }
\def\pmbh#1#2{\setbox0=\hbox{#1}%
    \setbox1=\hbox{#2}%
    \kern-.025em\copy0\kern-\wd0
    \kern.05em\copy1\kern-\wd0
    \kern-.025em\raise.0433em\box0 }

\def\frac#1#2{{#1\over#2}}

\def\boxit#1{\vbox{\hrule\hbox{\vrule\kern6pt
   \vbox{\kern6pt#1\kern6pt}\kern6pt\vrule}\hrule}}

\def\listing#1{\vskip 4mm\begin{verbatim}\input#1 \vskip 4mm}
\def\thick#1{\hbox{\rlap{$#1$}\kern0.25pt\rlap{$#1$}\kern0.25pt$#1$}}

\def\pmbh{{\pmb h}}

\def\calH{{\cal H}}

\def\calL{{\cal L}}

\renewcommand\today{\ifcase\month\or
   Jan\or Feb\or Mar\or Apr\or May\or
   Jun\or Jul\or Aug\or Sep\or Oct\or Nov\or
   Dec\fi
   \space\number\day, \number\year}


%% file: self-define-Heng.tex
\newcommand{\va}{{\bf a}}

\newcommand{\vx}{{\bf x}}

\newcommand{\vDelta}{\mbox{\boldmath $\Delta$}}
\newcommand{\vdelta}{\mbox{\boldmath $\delta$}}

\newcommand{\vbeta}{\mbox{\boldmath $\beta$}}

\newcommand{\vtheta}{\mbox{\boldmath $\theta$}}

\newcommand{\bay}{\begin{array}}
\newcommand{\eay}{\end{array}}

\newcommand{\bqa}{\begin{eqnarray*}}
\newcommand{\eqa}{\end{eqnarray*}}
\newcommand{\bqan}{\begin{eqnarray}}
\newcommand{\eqan}{\end{eqnarray}}
\newcommand{\bqt}{\begin{quote}}
\newcommand{\eqt}{\end{quote}}
\newcommand{\bt}{\begin{tabbing}}
\newcommand{\et}{\end{tabbing}}
\newcommand{\bit}{\begin{itemize}}
\newcommand{\eit}{\end{itemize}}
\newcommand{\ben}{\begin{enumerate}}
\newcommand{\een}{\end{enumerate}}
\newcommand{\beq}{\begin{equation}}
\newcommand{\eeq}{\end{equation}}
\newcommand{\bdefi}{\begin{definition}}
\newcommand{\edefi}{\end{definition}}
\newcommand{\bpro}{\begin{proposition}}
\newcommand{\epro}{\end{proposition}}
\newcommand{\bco}{\begin{corollary}}
\newcommand{\eco}{\end{corollary}}
\newcommand{\bdes}{\begin{description}}
\newcommand{\edes}{\end{description}}

